\documentclass[useAMS,usenatbib]{mn2e}
\usepackage[dvips]{graphicx}
\usepackage{amsmath}
\usepackage{amsfonts}
\usepackage{amssymb}

\title[Detection of a transit by the planetary companion of HD~80606]{Detection of a transit by the planetary companion of HD~80606}
\author[Stephen J. Fossey, Ingo P. Waldmann \& David M. Kipping]{Stephen J. Fossey$^{1}$\thanks{E-mail: sjf@ulo.ucl.ac.uk}\footnotemark[1], Ingo P. Waldmann$^{1}$ \& David M. Kipping$^{1}$\\
$^{1}$Department of Physics and Astronomy, University College London, 
Gower Street, London WC1E 6BT, UK}
\begin{document}

\date{Accepted 2009 March 10th. Received 2009 February 26th; in original form 2009 February 26th}

\pagerange{\pageref{firstpage}--\pageref{lastpage}} \pubyear{2009}

\maketitle

\label{firstpage}

\begin{abstract}
We report the detection of a transit egress by the $\sim 3.9$-Jupiter-mass planet HD 80606b, an object 
in a highly-eccentric orbit ($e\sim 0.93$) about its parent star of approximately solar type. The astrophysical reality of the signal of variability in HD 80606 is confirmed by observation with two 
independent telescope systems, and checks against several reference stars in the field. Differential photometry with respect to the nearby comparison star HD 80607 provides a precise light curve. Modelling of the light curve with a full eccentric-orbit model indicates a planet/star-radius ratio of $0.1057 \pm 0.0018$, corresponding to a planet radius of 1.029\,$R_J$ for a solar-radius parent star; and a precise orbital inclination of $89.285^\circ \pm 0.023^\circ$, giving a total transit duration of $12.1 \pm 0.4$ hours. The planet hence joins HD 17156b in a class of highly eccentric transiting planets, in which HD 80606b has both the longest period and most eccentric orbit. The recently reported discovery of a secondary eclipse of HD 80606b by the Spitzer Space Observatory permits a combined analysis with the mid-time of primary transit in which the orbital parameters of the system can be tightly constrained. We derive a transit ephemeris of $T_{\mathrm{tr}} = \mathrm{HJD} \,(2454876.344 \pm 0.011) + (111.4277 \pm 0.0032) \times E$.
\end{abstract}


\section{Introduction}

The `hot Jupiter' planetary companion to HD 80606, discovered through the reflex radial-velocity (RV) signature of the parent star by \citet{nae01}, is notable for the high eccentricity of its orbit at $e \sim 0.93$. The existence of such a system provides a laboratory for testing models of planetary formation and evolution (e.g., \citet{wu03} and \citet{for08}), particularly when the parent star is a member of a visual binary star, as here: its companion, HD 80607, lies at a separation of $21.1''$ (e.g., \citet{domm02}), corresponding to $\sim1000$ AU at the distance of about 60 pc (\citet{lau09}, hereafter L09).

The recent observation of a secondary eclipse of HD80606b by L09 with the Spitzer Space Observatory gives insight into the effects on the planetary atmosphere of the highly variable insolation of radiation from the parent star as the planet undergoes changes in irradiation of $\sim 800$ fold.

The importance of transits for characterizing the physical and atmospheric properties of exoplanets has been well documented (e.g. see \citet{win08}). Given the remarkable nature of the orbit and irradiation of HD 80606b, a transit observation would be especially valuable.  We reiterate here that transits determine the orbital inclination (and hence true planetary mass), radius and density, and thus constrain internal planetary structure (see \citet{zen08}).  They also provide opportunities for sensitive observation of the planetary atmospheric composition through IR absorption-line studies of the transit (e.g. \citet{tin07}), and emission properties during secondary eclipse (e.g. \citet{swa09}). Further, repeated, accurate measurement of the key parameters of the transit signal have the potential to probe more characteristics of the host planetary systems through a search for timing variations caused by resonant planets (\citet{ago05}; \citet{hol05}), satellites (\citet{kip09}) and Kozai migration (\citet{wu03}).  Hence, the reliable detection of a transit of HD80606b, combined with a timing of the secondary eclipse, would open up a very rich lode for investigation of the planet and host system.

Following their detection of a secondary eclipse at 8$\mu m$ with the IRAC instrument with Spitzer, L09 predicted the occurrence of a transit of HD 80606b with {\em a-priori\/} probability of $\sim$15\%, according to the ephemeris (L09) $T_{\rm tr} = {\rm HJD} 2454653.68 + 111.4277 \times N$.  The predicted event of ${\rm HJD}=2454876.5345$ (2009 February 14, 0050 UT) was well placed for observation from the UK, the object's declination meaning it would cross the meridian close to the zenith, and be available all night for continuous monitoring. Given the expected duration of a central transit of about 17 hours, and the uncertainty on the ephemeris timing, a global campaign to search for the transit ingress and egress signals was launched by the Transitsearch.org network.\footnote{See oklo.org}

In this paper, we describe a search for a predicted transit signal by HD 80606b using the facilities of the University of London Observatory (ULO). Our observations with two independent instruments show a clear signal which we interpret to be a transit egress.

\section{Observations}

The University of London Observatory, operated by UCL, is sited in Mill Hill, NW London. Observations of known and candidate exoplanet transiting systems have been carried out for several years, and in particular have involved UCL undergraduate students in a campaign to monitor targets of specific
interest. 

Several undergraduate members of the campaign were recruited to assist with telescope operations, and two instruments were used to monitor the field of HD 80606 throughout the night.

\begin{table*}
\caption{\emph{Properties of HD 80606 and photometric reference stars}} 
\centering 
\begin{tabular}{c c c c c} 
\hline\hline 
Star ID & Sp$^a$ & $V^b$ & $B-V^b$ & Comparison \\ [0.5ex] 
\hline 
HD 80606        & G5 & $9.06 \pm 0.04$  & $0.765 \pm 0.025$ & \\
HD 80607        & G5 & $9.17 \pm 0.04$  & $0.828 \pm 0.029$ & Ref.~1 \\
HD 233625       & G5 & $9.52 \pm 0.04$  & $0.62  \pm 0.04$  & Ref.~2 \\
TYC 3431-0892-1 &    & $10.04 \pm 0.04$ & $1.32  \pm 0.09$  & Ref.~3 \\ [1ex]
\hline\hline 
\end{tabular}
\label{table:stars} \\
$(a)$ Simbad CDS, {\tt http://simbad.u-strasbg.fr}\\
$(b)$ Hipparcos (\citet{esa97}) and Tycho-2 (\citet{hog00}), using the transformations of \citet{mama02}.
\end{table*}

\begin{enumerate}
\item Our primary observing programme used a Celestron 35-cm Schmidt-Cassegrain telescope on a Bisque Paramount ME German equatorial mount, with an SBIG STL-6303E CCD camera. The $22.2'\times 14.8'$ field was centred to include several bright reference stars on the chip 
(Table~\ref{table:stars}). An $R$-filter exposure time of 10 seconds was used for the entire series, yielding a cadence of about 20s, after readout overheads. Observations of the field were acquired from 1809--0414 UT, and 1638 useable science frames were obtained.
\item We operated a backup observing programme on a fork-mounted 25-cm Meade LX200 Schmidt-Cassegrain telescope, with an SBIG ST8-XME CCD camera providing a field of $18.6'\times 12.6'$.
Exposure times of 30 seconds were used, with a typical cadence of 35 seconds. With no requirement for the mount to be reversed as the target crossed the local meridian, these data provide an important check for the integrity of the Celestron photometry. A total of 761 useable science frames were obtained through a red filter ($\lambda_c\sim 650$\,nm, $\Delta\lambda\sim 100$\,nm) between 1833 and 0342 UT.
\end{enumerate}

Weather conditions were excellent throughout the night, and the atmospheric seeing was stable at about $2.5$--$3''$. The arrival of cloud shortly after 0400 UT forced all observations to end.

We used standard IRAF procedures for trimming, bias and dark subtraction, and flat-fielding. The bias and dark frames were obtained from a median stack of 10--20 frames, and the flat field from a median stack of 10 exposures of the evening twilight sky (Celestron) or dome (Meade).

\begin{figure}
\begin{center}
\includegraphics[width=8.4 cm]{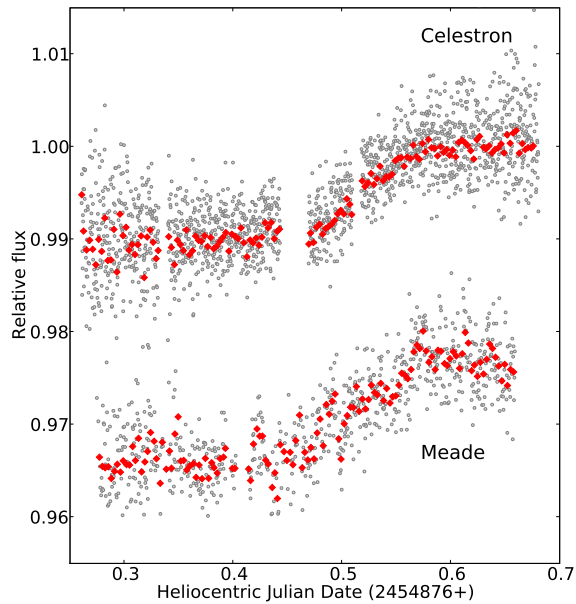}
\caption{\emph{The flux of HD 80606 relative to HD 80607. The relative fluxes have been normalized to unity using the mean of all data points at ${\rm HJD} > 2454876.575$; the Meade light curve has been
shifted vertically by $-0.023$ units for clarity. The unbinned light curves are plotted as grey filled symbols, while 4-minute binned data are plotted in red. The gaps in the Celestron light curve at $\sim .45$ and $\sim .51$ HJD are due to a system crash and the reversal of the mount respectively.}} \label{fig:fig1}
\end{center}
\end{figure}

\begin{figure}
\begin{center}
\includegraphics[width=8.4 cm]{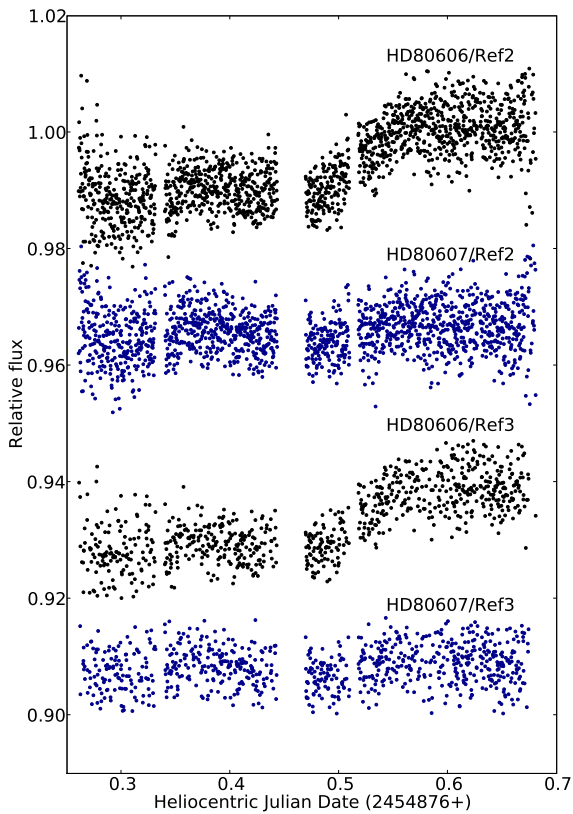}
\caption{\emph{The flux of HD 80606 and HD 80607 relative to comparison stars Ref. 2 and Ref. 3, from the Celestron photometry. The relative fluxes in each case have been normalized to unity using the mean of all data points at ${\rm HJD} > 2454876.575$; the lower light curves have then been shifted vertically for clarity.}} \label{fig:fig2}
\end{center}
\end{figure}

\section{Differential photometry}

We performed aperture photometry on all images using IRAF/DAOPHOT routines (\citet{stet87}). A tight star-aperture of radius 6 pixels ($\sim 5''$) and a sky annulus of inner radius $43''$ were used to avoid risk of contamination of HD 80606 by HD 80607, and {\em vice versa}, due to their proximity.

Fig.~\ref{fig:fig1} shows the flux ratio of HD 80606 relative to the nearby comparison HD 80607, for both instruments. In each case, the differential light curves have been renormalized to a flux ratio of unity using the mean of all data points later than $2454876.575$ HJD.

The light curves show a clear signal of variation, of order 1\%. Its appearance in the two independent datasets supports the reality of the signal: it is not an instrumental artefact. No attempt has been made to correct the light-curve for airmass-dependent effects; the observed variation is unlikely to be due to local, second-order, colour-dependent extinction effects, as the spectral types and colours of the two stars are very similar.

As an additional check on the reality of the signal, we checked the flux ratios of both stars against the other comparison stars in the Celestron field. The light curves are shown in Fig.~\ref{fig:fig2}, normalized in the same way as in Fig.~\ref{fig:fig1}, and show a consistent variation in HD 80606 relative to all comparison stars. The light curves for HD 80607 are, by contrast, relatively flat. There does appear to be a low-level residual structure in the HD 80607 light curves, but we believe these are likely to be due to flat-field residuals caused by shifts of the field registration on the chip, supported by their apparent correlation with residuals in HD 80606 also.

From the above, we confirm that the structure in the light curve seen in Fig.~\ref{fig:fig1} is due to variability in HD 80606. There are statistical advantages to combining the fluxes of all comparison stars to perform ensemble differential photometry with higher precision in each data point from the increased total flux and the averaging of scintillation noise; however, the apparent residual systematic effects seen in Fig.~\ref{fig:fig2} lead us to choose HD 80607 as the preferred comparison star for HD 80606.

\section{Light-curve analysis}
\subsection{Fitting}

The light curves for HD 80606 illustrated in Figs.~\ref{fig:fig1} and~\ref{fig:fig2} are highly reminiscent of a transit-egress signal by a Jupiter-sized companion (e.g., see the case of HD 17156b, \citet{barb07}). In order to follow-up this hypothesis, we modelled theoretical transit light-curves using the orbital parameters of HD 80606b given by L09, a stellar mass of $1 M_{\odot}$ (L09), and a planet mass of $3.9\,M_J$ (\citet{nae01}), and fitted them to the light curves presented in Fig~\ref{fig:fig1}. To account for the effects of high eccentricity, we use the dynamical model of \citet{kip08} (herafter, K08), coupled with the routines of \citet{man02} to produce limb-darkened lightcurves. The K08 model allows fully for the planet's velocity to vary during the transit and for light-curve asymmetry. 

We combined our data into 4-minute bins, and used the AMOEBA $\chi^2$-minimization routine (\cite{press92}) to find the best-fitting light curve to the binned data. We assumed a solar radius for HD 80606, after L09, and chose quadratic limb-darkening coefficients in the $R$-band from \citet{cla00} for $T_{\rm eff}=5500$ K, $\log g=4.50$ and $[{\rm Fe}/{\rm H}] = 0.5$, based on the stellar properties given by \citet{nae01}.

We performed fits to three data samples: $(i)$ Celestron only, $(ii)$ Meade only, and  $(iii)$ both datasets simultaneously. For $(i)$ and $(ii)$, the fits to each sample were unweighted. Investigation of the rms scatter of the residuals about each fit indicated that in the combined data, the Meade observations should be weighted by 0.25 relative to the Celestron data. We estimated uncertainties on our fitted parameters by boostrap resampling of the light-curve residuals and refitting in 1000 Monte Carlo simulations. All results are given in Table~\ref{table:table2}.

The results of the Celestron and Meade fits are internally consistent. Excluding the noisier data obtained at higher airmass at ${\rm HJD}<\,.35$, the rms scatter of residuals from the combined fit is $0.80$~mmag (Celestron) and $1.66$~mmag (Meade). This scatter is reasonably consistent with the photometric uncertainties calculated from the total flux and scintillation noise on each star.

The combined-fit results indicate a precise planet/star radius ratio of $0.1057 \pm 0.0018$, indicating a Jupiter-sized companion for a stellar radius of 1~$R_{\odot}$. The inclination is tightly constrained at $89.285 \pm 0.023$ degrees, yielding a total transit duration, $T_{1,4}$, of about 12 hours. 

\begin{figure}
\begin{center}
\includegraphics[width=8.4 cm]{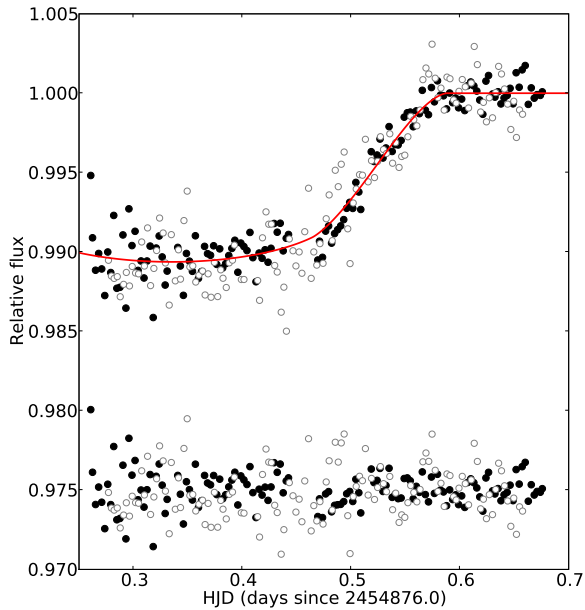}
\caption{\emph{The 4-minute binned data for the Celestron (filled symbols) and Meade (open) differential light curve of HD 80606 with respect to HD 80607, with the best-fitting eccentric-orbit light curve drawn through the points. The residuals from the best-fit curve are shown, at a relative flux of $+0.975$ units.}} \label{fig:fig3}
\end{center}
\end{figure}

\begin{table*}
\caption{\emph{Fitted planetary parameters}} 
\centering 
\begin{tabular}{l c c c} 
\hline\hline 
               & C-14 & Meade & Combined \\ [0.5ex] 
\hline
Number of binned data      & 139 & 134 & 273 \\
Central transit depth (\%) & 1.048 & 1.116 & 1.059 \\
Ratio-of-radii, $k$        &  $0.1043 \pm 0.0017 $ & $0.1100 \pm 0.0034$ & $0.1057 \pm 0.0018$\\
Egress duration, $T_{3,4}$ (hours) & $2.84\pm 0.22$ & $3.3\pm0.8$ & $2.98 \pm 0.26$ \\
Total duration, $T_{1,4}$ (hours)  &$12.3\pm 0.4$ & $11.9\pm 0.6$ & $12.1 \pm 0.4$ \\
Mid-transit time, $T_{\mathrm{mid}}$ & $0.339\pm 0.011 $ & $0.349\pm 0.018$ & $0.344 \pm 0.011$\\
\multicolumn{1}{r}{$({\rm HJD} - 2454876)$} & & & \\
Orbital inclination, $i$ ($^\circ$)& $89.296 \pm 0.032$ & $89.27 \pm 0.05$ & $89.285\pm 0.023$\\
Planetary radius, $R_P$ ($R_J$) & $1.015 \pm 0.016$ & $1.070\pm 0.030$ & $1.029 \pm 0.017$\\ [1ex]
\hline\hline 
\end{tabular}
\label{table:table2} 
\end{table*}

\subsection{Consistency of the signal with the predicted transit of HD 80606b}

Since the transit-like signal occurs in both data sets, we conclude it is of astrophysical origin. HD~80606 itself is not expected to be variable: indeed, the chromospheric activity of the star as measured by its Ca\,{\sc ii} H and K core emission (\citet{saffe05}) is indicated to be quiet. 

We can show that the mid-transit time of our best-fit model is consistent with the ephemeris of L09. Using the model of K08, we calculate the difference between the primary and secondary eclipse mid-points. The orbit parameters and uncertainties of L09 then give a predicted mid-transit of $T_{\mathrm{mid}} = 2454876.44 \pm 0.24$ HJD, compared to our observation of $2454876.344 \pm 0.011$ HJD, $\sim$ 2.5 hours early but well within the predicted uncertainty. The observed egress duration and depth constrains the second contact to occur at $\sim$ 1650 UT on 2009 February 13th, consistent with the absence of an observed ingress. Further, we can use the K08 model with our best fit parameters to predict the secondary-eclipse duration.  Our model predicts $s_{1,4} = 1.93$ hours and $s_{2,3} = 1.56$ hours, completely consistent with the results of L09 who found $s_{1,4} = 1.92 \pm 0.10 $ and $s_{2,3} = 1.68 \pm 0.24$ hours. 

We therefore conclude that there is a very high probability that the observed signal is indeed a planetary transit of HD80606b.

\begin{figure}
\begin{center}
\includegraphics[width=8.4 cm]{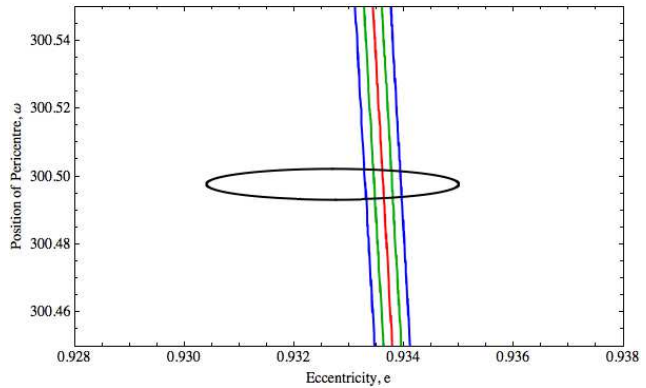}
\caption{\emph{The vertical lines represent the best-fit, one-sigma and two-sigma confidence limits on $e$ and $\varpi$.  The black ellipse represents the previously derived one-sigma constraints from L09. The two limits combined provide a highly precise measurement of $e$.
}} \label{fig:fig4}
\end{center}
\end{figure}

\subsection{Eccentricity constraints}

Based on the timing between the secondary and primary transit, it is possible to derive precise constraints on the orbital eccentricity.  L09 cited that the primary transit occurs only $\sim6$ days after the secondary eclipse, short in relation to the 111-day period due to the extremely high eccentricity. The eccentricity has been derived from RV measurements, and the time delay between primary and secondary eclipse can be used to constrain it very tightly. Typically, previous authors, for example \citet{win05}, have used a first-order approximation of $e$ for such a calculation to derive $e \cos \varpi$, but in this case we are very far from a `low' value of $e$. We have therefore used the K08 model to vary $e$ and $\varpi$ such that it reproduces the observed delay of $5.897 \pm 0.017$ days and we plot the allowed parameter values in Fig.~ 4.

Our constraints on $e$ and $\varpi$ can be seen to intersect the 1-sigma parameter space of L09 derived from independent RV measurements.  Combining both constraints, we derive highly precise values of $e = 0.9336 \pm 0.0002$ and $\varpi = 300.4977 \pm 0.0045$, or a precision in the eccentricity of 2 parts in 10000.

\section{Discussion and Conclusions}

We have detected a transit-egress signature towards HD 80606 in two sets of photometry taken simultaneously on the night of February 13/14 2009 from the ULO, London, UK.  The detection of the signal in two independent sets of photometry strongly supports an astrophysical origin for the signal.  Further, the moment of mid-transit is consistent with the ephemeris of L09 as well the egress duration being consistent with both the absence of observed ingress and the measured secondary-eclipse durations of L09.  We conclude that the signal is highly likely to be that of the transit of HD 80606b.

The planetary radius of 1.03 $R_J$, for a stellar radius of $1R_{\odot}$ implies a planet density of $4440 \pm 240 \,\mathrm{kg} \, \mathrm{m}^{-3}$, surpassing the density of the other known highly eccentric transiting planet, HD 17156b, and making HD 80606b an object of great interest to the field of planetary formation.

Hence, HD~80606b has the longest period and most eccentric orbit amongst the known transiting planets, making it an extremely unusual and interesting object for further study. The large change in stellar irradiation between primary and secondary eclipse offers an excellent opportunity for follow-up atmospheric studies.

We note that very long transit duration allows for extremely precise measurements of mid-transit time, depth and inclination.  Therefore, any periodic or secular modification to these parameters due to a perturbing body would be easier to detect relative to other systems. The planet's orbit is much wider than other transiting planets, and taking into account a significant light-time travel effect of 2.75 minutes between secondary and primary eclipse, we have demonstrated that the eccentricity of this system can be determined to a level of 2 parts in 10000. We derive a transit ephemeris of $T_{\mathrm{tr}} = \mathrm{HJD} \,(2454876.344 \pm 0.011) + (111.4277 \pm 0.0032) \times E$. 

Finally, the work reported here continues to highlight the utility of co-ordinated global campaigns which promote the use of relatively modest telescope apertures to observe transits and help characterize
exoplanet systems; indeed, such campaigns have been actively encouraged through the oklo.org website of G. Laughlin. The full characterization of the transit signal of HD 80606b will almost certainly depend on future co-ordination of observations made from multiple sites. We also emphasize the opportunities for useful work to be carried out from observing sites in locations --- even London --- which are not usually considered to be suitable for modern observational astrophysics. While there are bright stars hosting known or candidate transiting planets awaiting detection or characterization, this will remain true. 

[We note that at the time this paper was submitted, two independent reports of a detection of this same transit event were announced --- see Moutou et al. (arXiv:0902.4457) and Garcia-Melendo \& McCullough (arXiv:0902.4493).]

\section*{Acknowledgments}

The authors would like to extend special gratitude to Greg Laughlin for inspiring this attempt through his oklo.org website. We would like especially to thank the undergraduate members of the UCL 
team who assisted with the observations: Maria Duffy, Stephen Fawcett, Yilmaz Gul, and Cherry Ng.  
We thank Ian Howarth and Mike Dworetsky for discussions, advice and encouragement. SJF would like to thank Mick Pearson, Theo Schlichter and Peter Thomas for unflagging technical support at ULO, and Dan Smith and Yudish Ramanjooloo for their early development work on transit-monitoring from ULO. SJF thanks the UCL ESCILTA fund and the Royal Astronomical Society for financial support to facilitate student participation in the observing campaign. DMK is supported by STFC and UCL. This paper has made use of the SIMBAD database, operated at CDS, Strasbourg, France.

\label{lastpage}


\begin{thebibliography}{99}
\bibitem[\protect\citeauthoryear{Agol et al.}{2005}]{ago05} Agol, E., Steffen, J., Sari, R. \& Clarkson, W., 2005, MNRAS, 359, 567
\bibitem[\protect\citeauthoryear{Barbieri et al.}{2007}]{barb07} Barbieri, M., Alonso, R., Laughlin, G. et al. 2007, A\&A, 476, L13
\bibitem[\protect\citeauthoryear{Claret}{2000}]{cla00} Claret, A. 2000, A\&A, 363, 1081
\bibitem[\protect\citeauthoryear{Dommanget \& Nys}{2002}]{domm02} Dommanget, J. \& Nys, O. 2002, {\em Observations et Travaux}, 54, 5
\bibitem[\protect\citeauthoryear{ESA}{1997}]{esa97} ESA, {\em The Hipparcos and Tycho catalogues}, 1997, ESA SP-1200
\bibitem[\protect\citeauthoryear{Ford \& Rasio}{2008}]{for08} Ford, E. B. \& Rasio, F. A. 2008, ApJ, 686, 621
\bibitem[\protect\citeauthoryear{Hog et al.}{2000}]{hog00} Hog, E., Fabricius, C., Makarov, V. V. et al. 2000, A\&A, 355, L27
\bibitem[\protect\citeauthoryear{Holman \& Murray}{2005}]{hol05} Holman, M. J. \& Murray, N. W., 2005, Science, 307, 1288
\bibitem[\protect\citeauthoryear{Kipping}{2008}]{kip08} Kipping, D. M., 2008, MNRAS, 389, 1383 (K08)
\bibitem[\protect\citeauthoryear{Kipping}{2009}]{kip09} Kipping, D. M., 2009, MNRAS, 392, 181 
\bibitem[\protect\citeauthoryear{Laughlin et al.}{2009}]{lau09} Laughlin, G., Deming, D., Langton, J., Kasen, D., Vogt, S., Butler, P., Rivera, E. \& Meschiari, S. 2009, Nature, 457, 562 (L09)
\bibitem[\protect\citeauthoryear{Mamajek, Meyer \& Leibert}{2002,2006}]{mama02} Mamajek, E. E., Meyer, M. R. \& Liebert, J. L. 2002, AJ, 124, 1670 \& 2006, AJ, 131, 2360 
\bibitem[\protect\citeauthoryear{Mandel \& Agol}{2002}]{man02} Mandel, K. \& Agol, E. 2002, ApJ, 580, L171
\bibitem[\protect\citeauthoryear{Naef et al.}{2001}]{nae01} Naef, D. et al., 2001, A\&A, 375, L27
\bibitem[\protect\citeauthoryear{Press et al.}{1992}]{press92} Press, W. H. et al. 1992, {\em Numerical Recipes in FORTRAN77}, CUP
\bibitem[\protect\citeauthoryear{Saffe, G\'{o}mez \& Chavero}{2005}]{saffe05} Saffe, C., G\'{o}mez, M. \& Chavero, C. 2005, A\&A, 443, 609
\bibitem[\protect\citeauthoryear{Stetson}{1987}]{stet87} Stetson, P. B. 1987, PASP, 99, 191
\bibitem[\protect\citeauthoryear{Swain et al.}{2009}]{swa09} Swain, M. R., Vashisht, G., Tinetti, G., Bouwman, J., Chen, P., Yung, Y., Deming, D. \& Deroo, P, 2009, ApJL, in press
\bibitem[\protect\citeauthoryear{Tinetti et al.}{2007}]{tin07} Tinetti, G. et al., 2007, Nature, 448, 163
\bibitem[\protect\citeauthoryear{Winn et al.}{2005}]{win05} Winn, J. N. et al., 2005, ApJ, 631, 1215
\bibitem[\protect\citeauthoryear{Winn}{2009}]{win08} Winn, J. N., 2009, \emph{Transiting Planets}, IAU Symp., 253, 99
\bibitem[\protect\citeauthoryear{Wu \& Murray}{2003}]{wu03} Wu, Y., Murray, N., 2003, ApJ, 589, 605
\bibitem[\protect\citeauthoryear{Zeng \& Seager}{2008}]{zen08} Zeng, Li \& Seager, S. 2008, PASP, 120, 871
\end{thebibliography}
\end{document}